\documentstyle[emulateapj,psfig]{article}

\begin{document}

\title{Implications of the Narrow Period Distribution of 
Anomalous X-ray Pulsars and Soft Gamma-Ray Repeaters}

\author{Dimitrios Psaltis\altaffilmark{1} and
M.\ Coleman Miller\altaffilmark{2}}

\altaffiltext{1}{School of Natural Sciences, Institute for Advanced
Study, Einstein Dr., Princeton, NJ 08540; dpsaltis@ias.edu}

\altaffiltext{2}{Department of Astronomy, University of Maryland,
College Park, MD 20742-2421; miller@astro.umd.edu}

\begin{abstract}
The spin periods of the ten observed Anomalous X-ray Pulsars and Soft
Gamma-ray Repeaters lie in the very narrow range $6-12$~s. We use a
point likelihood technique to assess the constraints this clustering
imposes on the birth period and on the final period of such systems.
We consider a general law for their spin evolution described by a
constant braking index. We find that, for positive values of the
braking index, the observed clustering requires an upper cut-off
period that is very close to the maximum observed period of $\simeq
12$~s. We also show that the constraint on the birth period depends
very strongly on the assumed value of the braking index $n$, ranging
from a few milliseconds for $n\gtrsim 2$ to a few seconds for
$n\lesssim 2$.  We discuss the possible ways of tightening these
constraints based on similarities with the population of radio pulsars
and with future observations of such sources with current X-ray
observatories.
\end{abstract}

\mbox{}


\section{INTRODUCTION}

In the last few years, evidence for the existence of neutron stars
with ultrastrong magnetic fields, or magnetars, has become very
compelling.  The discovery of rapid spin down in the pulsations
observed from soft gamma-ray repeaters (SGRs; e.g., Kouveliotou et
al.\ 1998, 1999) gave support to the suggestion by Thompson \& Duncan
(1995) that the very energetic bursts observed from these sources
require the presence of $10^{15}$~G magnetic fields.  Furthermore, the
lack of detectable companions to the Anomalous X-ray pulsars (AXPs;
e.g., Mereghetti, Israel, \& Stella 1998; Hulleman, van Kerkwijk, \&
Kulkarni 2000), combined with their rapid spin-down rates and spectral
properties (see, e.g., \"Ozel 2001), favor the magnetar
interpretation.

A striking feature of all magnetar candidates (SGRs and AXPs) is that
their periods lie in a relatively narrow range, between 6~s and 12~s.
This property has been discussed since the original study of
Mereghetti \& Stella (1995) and was addressed by Colpi, Geppert, \&
Page (2000) in the context of models with magnetic field decay (see
also Chatterjee \& Hernquist 2000 for a discussion of the period
clustering expected in a variant of accretion models for AXPs).
However, a quantitative, statistical analysis of the constraints
imposed on magnetar models by this period clustering is still lacking.

Here we quantify the tightness of the observed period distribution,
using a point likelihood technique.  In \S~2 we review the detections
of each of the 10 magnetar candidate sources and focus in particular
on selection effects that could restrict the range of periods that can
be discovered.  In \S~3 we perform a likelihood analysis to determine
the allowed range of periods using a mathematical model for the period
evolution of AXPs that has broad applicability and of which a dipole
spin-down law is a special case. In \S~4 we discuss the implications
of these results and explore the potential for improving these
constraints using future observations of AXPs and SGRs with current
X-ray telescopes.

\section{OBSERVATIONS AND SELECTION EFFECTS}

In this section we describe briefly the observations that led to the
discovery of the ten AXPs and SGRs with measured spin periods, in
order to assess the observational selection effects that could have
affected their period distribution. Even though no systematic searches
for AXPs and SGRs have been performed to date, we argue that these
discovery observations were not confined to periods of order $\sim
6-12$~s and hence no observational selection effects can account for
the observed period clustering.

\subsection{Anomalous X-ray Pulsars}

Two of the AXPs were known persistent X-ray sources before pulsations
were detected in their X-ray emission. 4U~0142$+$61 ($P=8.69$~s) was
discovered as a persistent source in an early all-sky
survey. Subsequent power-spectral analysis of {\em pointed\/}
observations of this source with {\em EXOSAT\/} revealed the
pulsations (Israel, Mereghetti, \& Stella 1994). The search was
performed using {EXOSAT ME\/} data that had a timing resolution of
1~s, yielding a lower limit on the search period of only 2~s. The
duration of the observation was 12~h and the upper limit on the search
period was of order $10^4$~s.

1E~1048.1$-$5937 ($P=6.44$~s) was serendipitously discovered as an X-ray
source with Einstein.  The pulsations were discovered by Seward,
Charles, \& Smale (1986), who searched in both Einstein and {\em EXOSAT\/}
data but did not report the period range of their searches.

The remaining three AXPs were discovered in searches for pulsed X-ray
sources, and hence the strategy followed in their observations could
have introduced significant selection effects.  1E~2259$+$586
($P=6.98$~s) was discovered using Einstein data, following a systematic
search for a pulsar in the supernova remnant G~109.1$-$1.0 (Fahlman \&
Gregory 1981). The range of periods searched was 0.1$-$200~s.

1E~1841$-$045 ($P=11.77$~s) was discovered with {\em ASCA\/},
following a systematic search for pulsations from all point sources
within the supernova remnant Kes~73 (Vasisht \& Gotthelf 1997). The
search was performed with three timing resolutions: 488~$\mu$s, 32~ms,
and 0.5~s and power spectra were produced for each timing
resolution. As a result, the minimum period at which pulsations could
be detected was $\ll 1$~s. Moreover, the search was performed over
timescales of 96~min, 10~min, and 1~min and thus was sensitive to
pulsations with periods $\gg 100$~s.

1RXS~J170849.0$-$400910 ($P=10.99$~s) was discovered by {\em ASCA\/}
in a survey of the galactic plane and was later identified with a {\em
ROSAT\/} source (Sugizaki et al.\ 1997). The observation was performed
with a timing resolution of 62.5~ms in the high-bit-rate mode and
0.5~s in the medium-bit-rate mode. As a result, searches for
pulsations were sensitive to periods $\ll 1$~s. The highest period
searched was quoted as $\simeq 600$~s.

AX~J1845$-$0258 ($P=6.97$~s) was discovered with {\em ASCA\/} in the distant
Milky Way (Gotthelf \& Vasisht 1998) in the supernova remnant
Kes~75. For computational efficiency, the search was performed only
for long periods, i.e., for 1~s$<$P$<$100~s.
 
\subsection{Soft Gamma-Ray Repeaters}

SGRs are identified through their recurrent $\gamma$-ray bursts and
not because of their pulsations. However, pulsations have been
detected in {\em all} four, securely identified SGRs. In
SGR~1900$+$14, pulsations have been observed both in the quiescent
emission and during bursts at a period of $P=5.16$~s (Hurley et al.\
1999).  In SGR~1806$-$20 ($P=7.47$~s) and SGR~1627$-$41 ($P=6.4$~s)
pulsations have been detected only during the quiescent emission
(Kouveliotou et al.\ 1998; Woods et al.\ 1999) whereas in
SGR~0525$-$66 ($P=8$~s) pulsations have been observed only during
bursts (Barat et al.\ 1979). All these searches were performed with
data obtained using {\em ASCA} or {\em RXTE} and, therefore, were not
limited to periods only comparable to those observed.

\section{ANALYSIS OF PERIOD CLUSTERING}

The period clustering of AXPs and SGRs has often been attributed to a
general prediction of a large class of spin-evolution models in which
the spin period derivative decreases with increasing period.  In these
models, the objects evolve quickly through the small periods, making
their detection improbable at these periods and their steady-state
period distribution insensitive to the birth values. However, because
the objects spend increasingly more time at long periods, the observed
cutoff towards high periods can be used for placing a very strong
constraint on the maximum period at which they are detectable. In this
section we quantify the above statement using a point likelihood
technique.

\subsection{Analytical Setup}

In order to model the period distribution of magnetar candidates, we
assume a general braking law of the form
\begin{equation}
 \dot{\Omega}=-\kappa \Omega^n\;,
 \label{eq:braking}
\end{equation}
where $\Omega$ is the spin frequency of the stars and $n$ is the
braking index. For $n=3$ and $\kappa\sim B^2$,
equation~(\ref{eq:braking}) corresponds to the standard spin-down law
for an inclined magnetic dipole of strength $B$. In order to avoid
introducing unnecessary complications to our model, we assume that all
systems are born with the same initial period $P_{\rm in}$ and become
undetectable when they reach period $P_{\rm f}$.

The evolution of the period distribution function $f(P)$ between
$P_{\rm in}$ and $P_{\rm f}$ is described by the conservation law
\begin{equation}
\frac{\partial f(P)}{\partial t}+\frac{\partial}{\partial P}
   \left[\dot{P}f(P)\right]=0\;,
\label{eq:pde}
\end{equation}
where we have assumed that there is no evolution of the factor $\kappa$.
In steady state, the distribution of systems over spin
period is then $f(P)\sim \dot{P}^{-1}$, or
\begin{equation}
f(P)=C P^{n-2}\;,
\end{equation}
where the constant $C$ is calculated from the requirement that $f(P)$ is
normalized, or
\begin{equation}
C=\left\{ \begin{array}{ll}
	  (n-1)(P_{\rm f}^{n-1}-P_{\rm in}^{n-1})^{-1}  & , n\ne 1\\
	  \ln\left(-P_{\rm f}/P_{\rm in}\right) & , n=1
          \end{array}
  \right.\;.
\end{equation}

We now use this period distribution to estimate the best values for
the initial and final period, using a likelihood analysis and given
the fact that $m$ systems have been detected with measured periods
$P_j$; $j=1,...,m$. (We assume that all periods are measured to
arbitrary precision.) To this end, we subdivide the available
parameter space into infinitesimally small bins of width $\Delta P$,
such that each bin has either zero or one data point in it.  The
likelihood of the data given the model is then simply
\begin{equation}
{\cal F}(P_{\rm j}\vert P_{\rm in}, P_{\rm f}) = 
   \prod_{j=1}^{m} f(P_j)\Delta P = C^m (\Delta P)^m 
   \prod_{j=1}^m P_j^{n-2}\;.
\end{equation}
If we assume two prior probability distributions ${\cal G}(P_{\rm
in})$ and ${\cal G}(P_{\rm f})$ for the parameters, then the posterior
probability distribution is proportional to the above
likelihood. Following the standard Bayesian approach, we then obtain
the posterior probability distribution for an individual parameter
(e.g., $P_{\rm in}$) by integrating the full multidimensional
posterior distribution over all parameters but the one of interest,
i.e., ``marginalizing'' over the remaining parameters. In practice,
the posterior distribution of, e.g., $P_{\rm in}$ is
\begin{equation}
{\cal P}(P_{\rm in}| P_{\rm j})=
\frac{\int {\cal F}(P_{\rm j}\vert P_{\rm in}, P_{\rm f}) 
          {\cal G}(P_{\rm in}){\cal G}(P_{\rm f}) dP_{\rm f}}
     {\int \int {\cal F}(P_{\rm j}\vert P_{\rm in}, P_{\rm f}) 
          {\cal G}(P_{\rm in}){\cal G}(P_{\rm f}) dP_{\rm f} dP_{\rm in}}\;.
\end{equation}

For the prior probability distribution over $P_{\rm in}$ we assume a
flat distribution in $\log P_{\rm in}$ between $10^{-3}$~s and the
minimum observed period of $P_{\rm min}=6.44$~s, which does not imply
any particular period scale. We chose $10^{-3}$~s as our lowest
acceptable initial spin period, because this is comparable to the
fastest neutron-star spins allowed by 

\centerline{\psfig{file=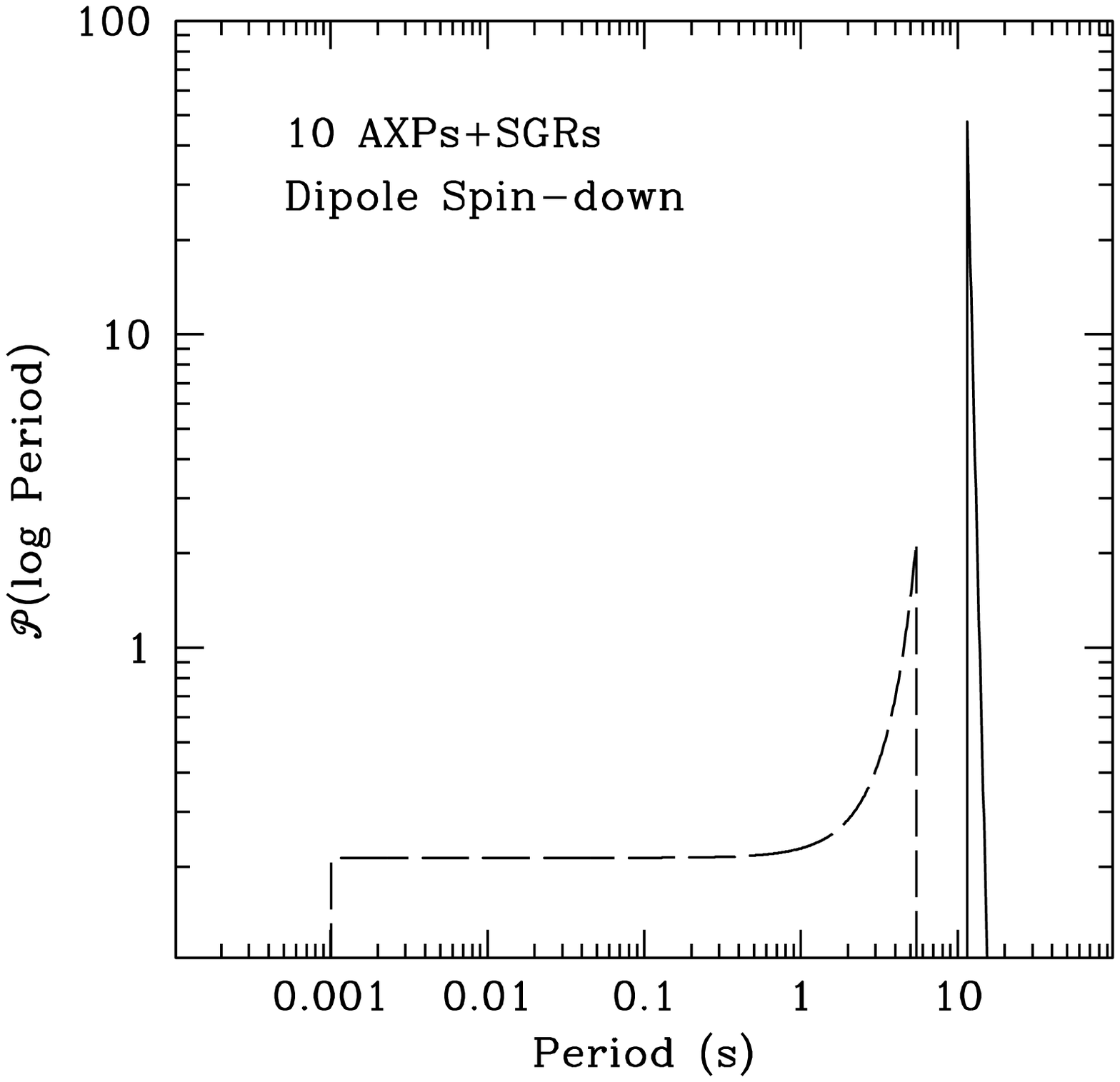,height=7.0truecm}}
\figcaption[]{\footnotesize The posterior probability distributions over
the initial (dashed line) and final (solid line) spin periods calculated
for the ten AXPs and SGRs discussed in \S2 and for a dipole spin-down law.}

\mbox{}

\noindent general relativity and modern equations of state (Cook, Shapiro \&
Teukolsky 1994). Similarly, for the prior probability distribution
over $P_{\rm f}$, we assume a flat distribution in $\log P_{\rm f}$
between the maximum observed period $P_{\rm max}=11.77$~s and
$100$~s. The upper acceptable final spin period is arbitrary and
affects very weakly the results for positive values of the braking
index.

Using equations~(3) and (4), we derive the posterior probability
distribution of, e.g., the initial period $P_{\rm in}$, to be
\begin{equation}
{\cal P}(P_{\rm in}) =
 \frac{ P_{\rm in}^{-1}\int_{P_{\rm f}} P_{\rm f}^{-1}
     (P_{\rm f}^{n-1}-P_{\rm in}^{n-1})^{-m} dP_{\rm f}}
      {\int_{\rm P_{\rm in}} P_{\rm in}^{-1}\int_{P_{\rm f}} P_{\rm f}^{-1}
     (P_{\rm f}^{n-1}-P_{\rm in}^{n-1})^{-m} dP_{\rm f} dP_{\rm in}}\;,
\end{equation}
where we have shown for simplicity only the expression for $n\ne
1$. Note that, because we are not testing the hypothesis that the
period distribution of sources follows a power law but we are simply
estimating parameters, the posterior distributions depend only on the
range of observed periods and not on their specific values.

\subsection{Numerical Results}

Figure~1 shows the posterior probability distributions over $\log
P_{\rm in}$ and $\log P_{\rm f}$ for a dipole spin-down law (i.e., for
$n=3$) and for the 10 magnetar candidates discussed in \S2. Clearly,
for both parameters, the most likely values are the extremes of the
observed period range. However, the shapes of the probability
distributions are very different for the two parameters.

For the dipole spin-down law assumed, the systems spend increasingly
more time at increasingly longer periods. Therefore, the absence of
any observed systems with periods larger than 12~s requires a rather
rapid turn off at periods comparable to the highest observed
period. As a result, the probability distribution over $\log P_{\rm
f}$ is very sharply peaked. On the other hand, for this spin-down law,
the initial period $P_{\rm in}$ is nearly unconstrained. Systems that
appear now at periods $\simeq 6-12$~s have spent very little time
slowing down from $\sim 10^{-3}$~s to $\sim 1$~s and, therefore, there
is little information about their initial 

\centerline{\psfig{file=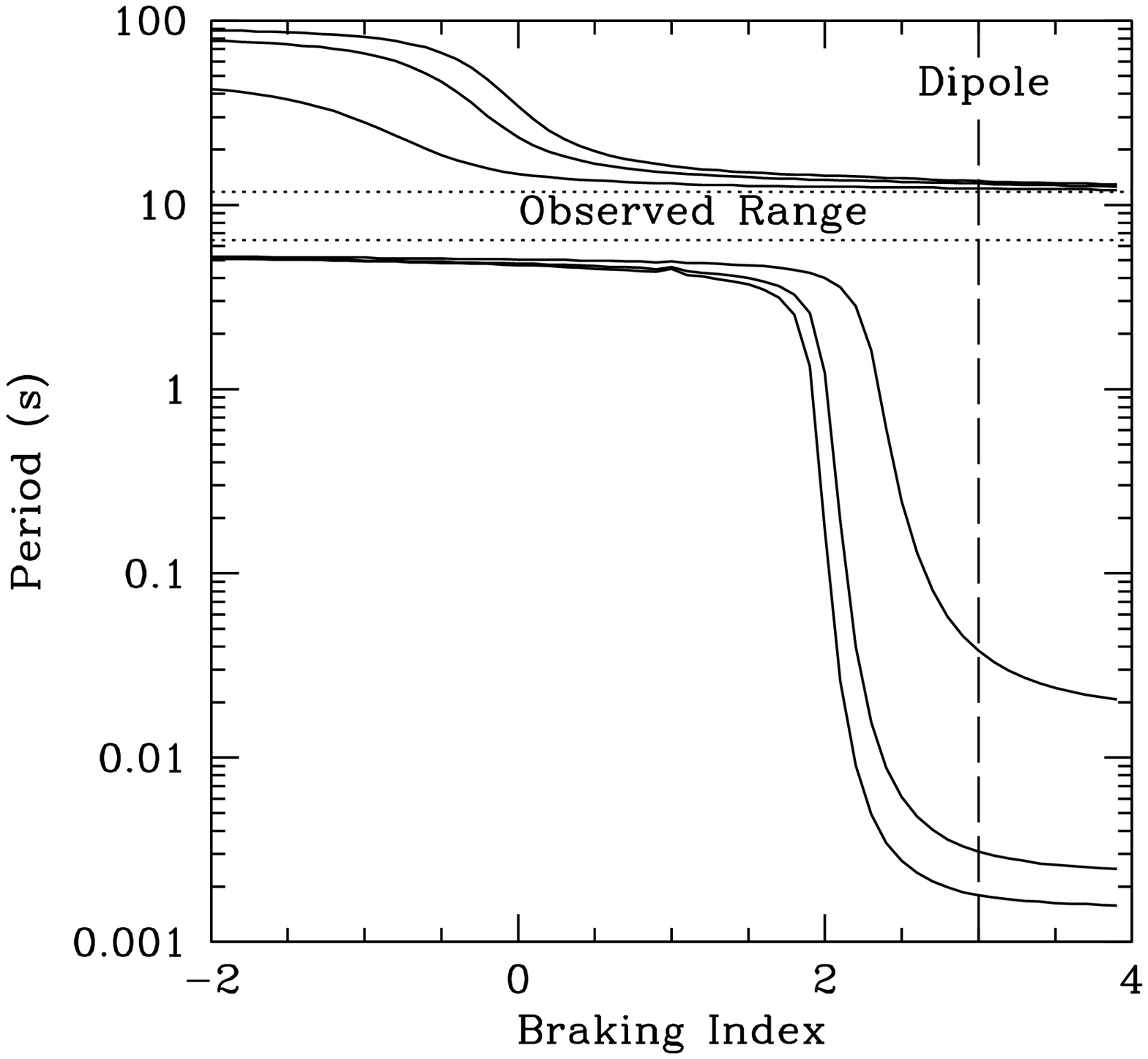,height=7.0truecm}}
\figcaption[]{\footnotesize The 68\%, 90\%, and 95\% confidence
levels of the initial (lower part of the diagram) and final (upper
part) periods of AXPs and SGRs for different values of the braking
index.}

\mbox{}

\noindent periods imprinted on their current period distribution.

As it is apparent from the above discussion, the constraints on the
initial and final periods of AXPs and SGRs depend strongly on the
braking index. This is shown in Figure~2, where the 68\%, 90\%, and
95\% confidence levels of $P_{\rm in}$ and $P_{\rm f}$ are plotted
against the assumed braking index $n$. For $n>2$ the final period is
strongly constrained to lie very close to the maximum observed period
while the initial period can lie in a large range of values. On the
other hand, the situation is reversed for $n<0$. In both cases, the
flattening of the confidence limits at the extreme values of allowed
periods is caused by the assumed prior distributions that are
bounded. This is a physical bound for the case of the initial period,
as discussed above, but it is artificial for the case of the final
period.

\section{DISCUSSION}

The narrow range of observed periods of magnetar candidates can be
used to constrain their birth periods, the periods at which they cease
to be active, or both, depending on the value of their braking
index. For positive values of the braking index, we showed in \S3 that
the final periods must lie very close to the maximum observed period
of $\simeq 12$~s. At the same time, if the braking index is $n\lesssim
2$ then the birth periods must be $\simeq 5$~s, i.e., very slow. On
the other hand, if $n\gtrsim 2$, then the birth periods are largely
unconstrained from the analysis of the observed clustering.

It is interesting to note that braking indices of young radio pulsars
have been measured to be from as low as 1.81$\pm$0.07 (PSR~B0540$-$69;
Zhang et al.\ 2001) to as high as 2.837$\pm$0.001 (PSR~B1509$-$58;
Kaspi et al.\ 1994), bracketing the value of $n=2$ that separates a
strong constraint on $P_{\rm in}$ from a very weak constraint.
However, slow radio pulsars have second period derivatives (and hence
braking indices) that are variable and larger by factors of $\sim
10^2-10^4$ than what is predicted by simple magnetic braking (see, e.g.,
Cordes \& Helfand 1980). Such anomalously large braking indices are
thought to be the result of glitches and of timing noise, both of
which are known to occur at least in AXPs, which can have

\centerline{\psfig{file=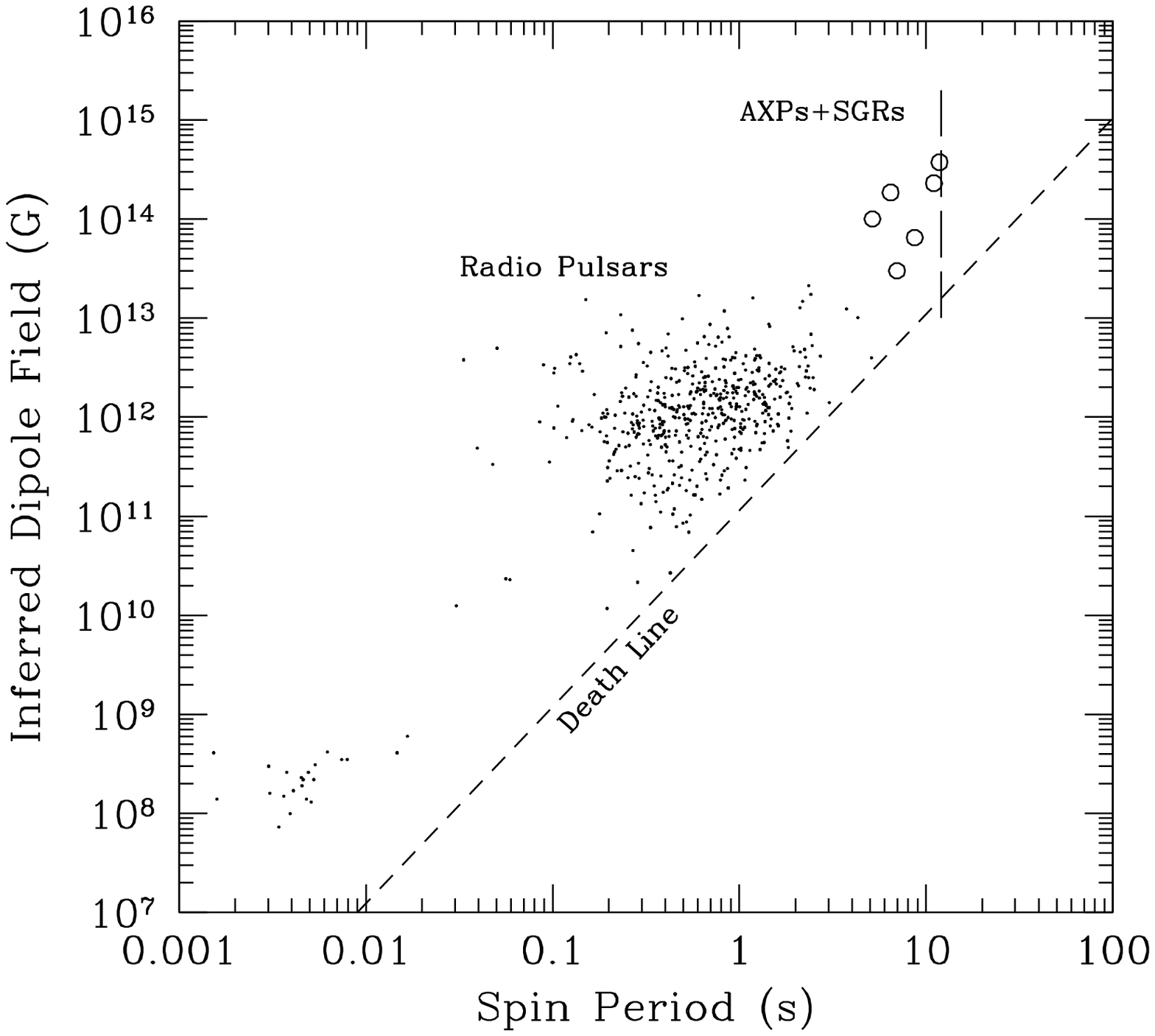,height=7.0truecm}}
\figcaption[]{\footnotesize The inferred {\em equatorial\/} dipole
magnetic fields of radio pulsars, AXPs, and SGRs versus their spin periods.
The short-dashed line shows the empirical death line for radio pulsars;
the long-dashed line shows the upper limit in the AXP and SGR periods
inferred in this paper.}

\mbox{}

\noindent instantaneous braking indices of order $\sim 10^3$ 
(Kaspi, Lackey, \& Chakrabarty 2000; Kaspi et al.\ 2001).

Our analysis, however, and in particular the constraint on the birth
periods, depends on the value of the average braking index, to the
extent that the spin evolution of the magnetar candidates is described
approximately by a law of the form~(\ref{eq:braking}), and not on any
instantaneous index. For the case of twenty moderate-aged radio
pulsars, Johnston \& Galloway (1999) computed braking indices
integrated over $\simeq 5-20$~yr and found values in the wide range
$-220\lesssim n\lesssim 35$. This result is not surprising, given that
the braking indices of such pulsars computed over shorter timescales
vary rapidly and often change sign. It shows, however, that
unfortunately little progress can be expected in constraining the
braking indices of magnetar candidates and hence their birth periods.

On the other hand, the strong constraint derived here on the maximum period 
may provide some insight into the physical mechanism that might
be causing it. For example, comparing the inferred {\em equatorial\/}
dipole magnetic fields and spin periods of magnetar candidates with
those of radio pulsars (see Fig.\ 3) indicates that the {\em
empirically drawn} death line of radio pulsars, when extrapolated
towards higher magnetic fields, appears to be unrelated to the maximum
period of magnetar candidates.

An exponentially decaying magnetic field, as discussed by Colpi et
al.\ (2000), provides a plausible explanation for both the period
clustering and the young ages of magnetar candidates. It is worth
noting, however, that neither the inferred luminosities of AXPs nor
their pulsed fraction 

\centerline{\psfig{file=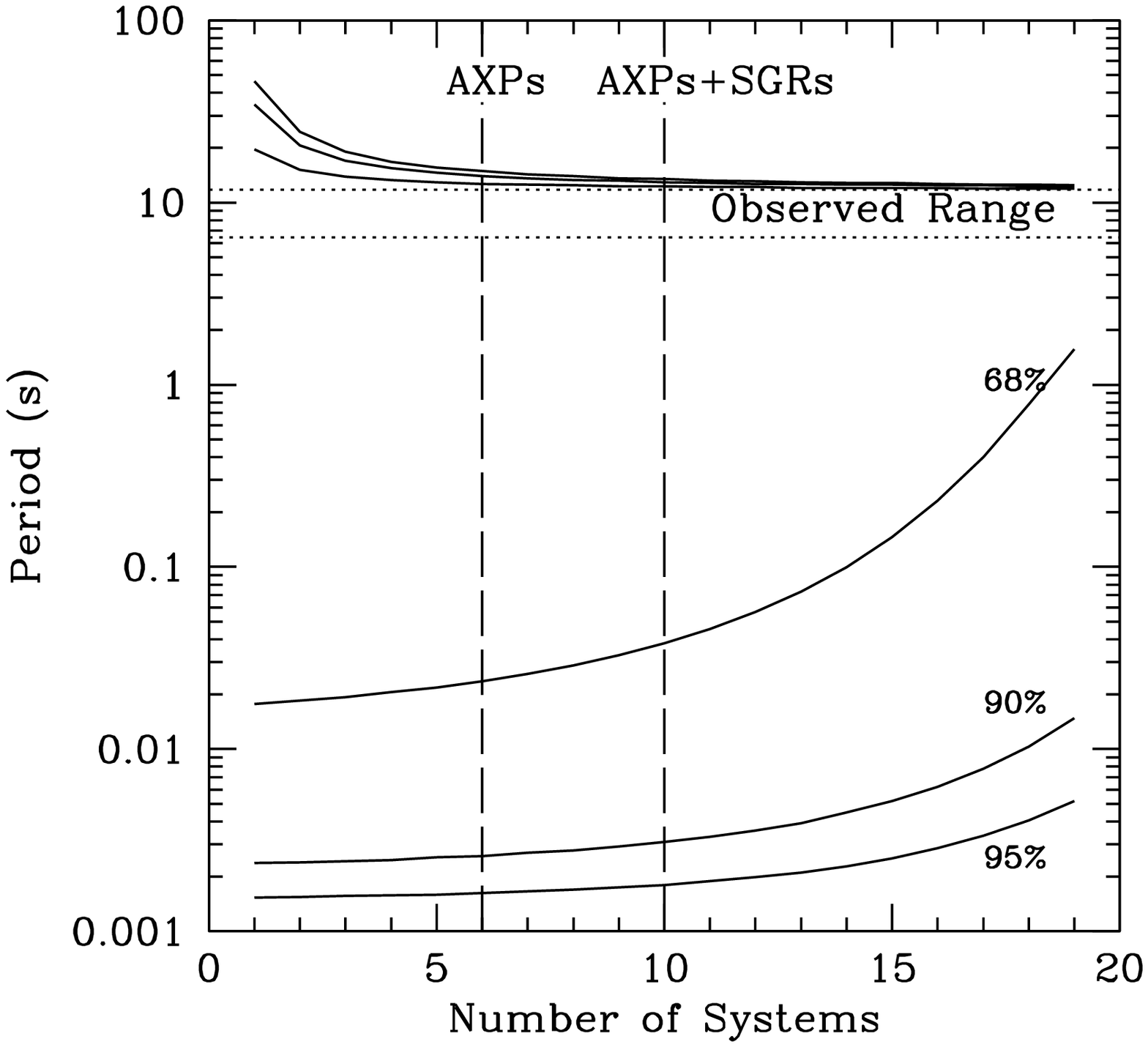,height=7.0truecm}}
\figcaption[]{\footnotesize The 68\%, 90\%, and 95\% confidence
levels of the initial (lower part of the diagram) and final (upper
part) periods of magnetar candidates, as a function of their total
number, for a dipole braking law.}

\mbox{}

\noindent appear to be correlated with their periods,
spin-down ages, or inferred field strengths, even though the values of
the latter two span more than an order of magnitude (see, e.g.,
\"Ozel, Psaltis, \& Kaspi 2001). 

Finally, we address the dependence of our results on the number of
magnetar candidates that we considered. In our analysis, we chose to
assume that both AXPs and SGRs are formed and evolve in the same way.
We show in Figure~4, however, where we use the dipole spin-down law as
an example, that the resulting constraints depend rather weakly on the
number of systems that are known within {\em the same\/} period
range. Indeed, even considering simply the six known AXPs would be
enough to reach similar conclusions.

Figure~4 also shows that increasing the sample of magnetar candidates
with similar periods even by a factor of two will only affect our
results mildly: the constraint on the birth period will become as high
as $\sim 2$~s, but only at the 68\% level.  However, the detection of
even a single magnetar candidate with period larger than $\simeq 12$~s
will change the constraint on $P_{\rm f}$. Such a detection does not
require detectors with fast timing capabilities and is, therefore,
possible, if such systems exist, with current X-ray observatories such
as Chandra and XMM/Newton.

\acknowledgements
 
We thank Feryal \"Ozel for many useful discussions and comments on the
manuscript.  D.\,P.\ thanks the Astronomy Department of the University
of Maryland for its hospitality. M.\,C.\,M.\ acknowledges the support
of NASA grant 5-9756 and NSF grant AST 0098436.

\end{document}